# SPEECH SEGMENTATION USING MULTILEVEL HYBRID FILTERS

*Marcos Faúndez Zanuy, Francesc Vallverdú Bayes*
Department of Signal Theory and Communications (UPC)
c/Gran Capità S/N   (Campus Nord, Edifici D5)
08034 BARCELONA (SPAIN)
tel:+34 3 4016453 fax:+34 3 4016447
e-mail:marcos@gps.tsc.upc.es

**ABSTRACT**

A novel approach for speech segmentation is proposed, based on Multilevel Hybrid (mean{}/min{}) Filters (MHF) with the following features:
• An accurate transition location.
• Good performance in noisy environments (gaussian & impulsive noise).

The proposed method is based on spectral changes, with the goal of segmenting the voice into homogeneous acoustic segments.

This algorithm is being used for phonetically-segmented speech coder, with successful results.

## 1. INTRODUCTION

Speech applications usually require to segment the speech signal into subword units that represent regions of quasi-stationarity in the spectral domain. Thus, each subword is composed of multiple highly correlated frames. Usually, the goal is to reduce the quantity of information to code, save or process.

Speech segmentation algorithms can operate with waveform and spectral parameters. The method that we propose is based on the second ones.

This paper is organized as follows: section two describes the algorithm. Section three deals the different parameters related with the algorithm and his influence over the final result. Section four compares the obtained results with the MHF with other methods. Finally, section five summarizes the main conlussions and results.

## 2. THE ALGORITHM

The algorithm that we propose for segmenting the speech signal can be summarized in the following steps:

Step 1. An analysis window of nine frames is selected:
$$([frame_{-4},...,frame_0,...,frame_4])$$
Figure 1 represents the frames used simultaneously. Frame zero represents the frame under test. If his left and right contexts are similar, then it will be declared stationary. Otherwise, there is a transition.

Step 2. A spectrum estimation of each frame is computed.

Step 3. multilevel hybrid filter is applied. It is named multilevel because the output of the filter is the result of two successive stages of filtering. Also it is hybrid because it combines linear and non linear filtering.

In the first stage of the MHF 8 averaged frames are obtained. We define four averages in each context (left and right). See fig. 1

$$\Phi_{ri} = \mathrm{mean}_{k=0,i}\{frame_k\}_{i=1,4} \quad \Phi_{li} = \mathrm{mean}_{k=-i,0}\{frame_k\}_{i=1,4}$$

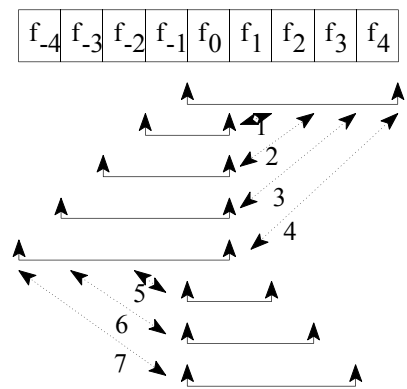

Figure 1. Frames averaged in each subwindow.

Step 4. Between these windows, we define seven differences (see fig. 1):

$$D_i = \text{error}(\Phi_{ri}, \Phi_{li})^2 \quad \text{if } i \neq 4$$
$$D_{i|4} = \text{error}(\Phi_{l4}, \Phi_{ri})^2 \quad \text{if } i \neq 3$$

obtaining $D = [D_1\ D_2\ D_3\ D_4\ D_5\ D_6\ D_7]$.

Step 5. In the second level of the MHF, the following nonlinear filter is applied to the output of the previous level:

$$\text{output} = \min\{D\}$$

If the max{} filter is chosen, there is an anticipative tendency to detect transitions, and they keep on being detected in posterior frames. This can be seen in fig 2.2. The peaks of the MHF output are wider than MHF output with min{} filter. Fig 2.1 shows very sharp peaks, which are easier to detect (position of transition frame) than peaks of figure 2.2. Obviously, the amplitude is greater in the second figure.

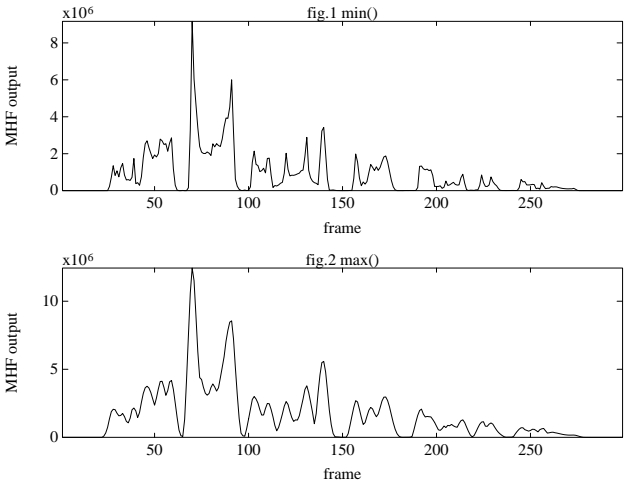

Figure 2. MHF output with min and max filters.

The final output of the MHF indicates the transition level of the window under test. This value must be compared with the neighbour outputs in order to decide if there is a transition or not.

Figure 3 shows the output of the MHF if an LPC analysis is realized instead of fft (results of figure 2 are obtained with the same sentence). Obviously the result is intrinsically independent of the frame energy, so no additional normalization is required.

## 3. PARAMETERS

The main selected parameters in our simulation are:

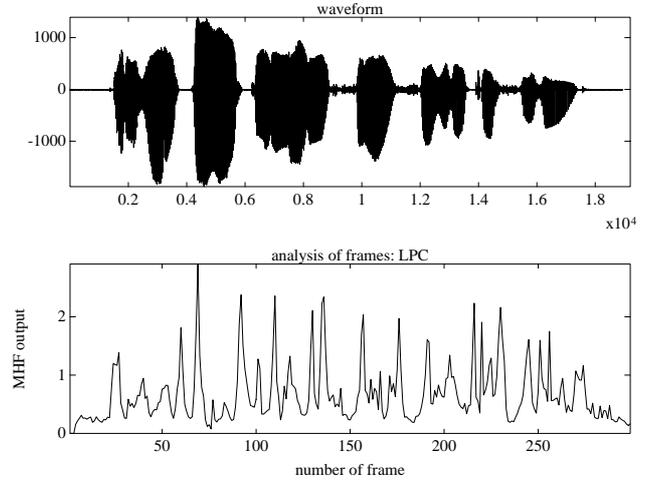

fig 3. MHF output with LPC analysis.

• Speech signal is sampled at 8KHz and 12 bits/sample.

• Experimentally we have fixed frame length to 16 ms and overlap to 50%. Higher lengths imply problems for resolving transitions with accuracy. Smaller frame lengths produce a bad spectral representation of the frames.

• Error measures studied are norms ($\|\cdot\|_1, \|\cdot\|_2, \|\cdot\|_\infty$), scalar product, Canberra, Tanimoto [4]. $\|\cdot\|_2$ produces good enough results, so it has been chosen.

• The best combination of filters that we have found is mean and min for levels 1 and 2 respectively. (Other filters tested are median, max, α-trimmed mean, etc. [5])

## 4. COMPARATIVE STUDY

Compared with similar measures [1], [3], the proposed method obtains a more accurate location of the transitions and immunity to noise because of the min{D} function. A transition only can be detected if it is indicated simultaneously by all the $D_i$. Preliminar results also indicate that the MHF is minimally affected if noise is added to the signal, because first stage produces a reduction of correlated noise, and second stage reduces impulsional noise.

Also, a very sharp energy change detector is obtained using the same MHF algorithm with the only change of the error measure by $v_1 * \text{pinv}(v_2)$.

## 5. RESULTS AND MAIN CONCLUSIONS

Figure 4 shows the percentage of use of each window. Left diagram is obtained with the mean/min MHF and right diagram with mean/max MHF. Obviously windows 1 and 5 are more often chosen for the min filter, and window 4 in the case of max filter.

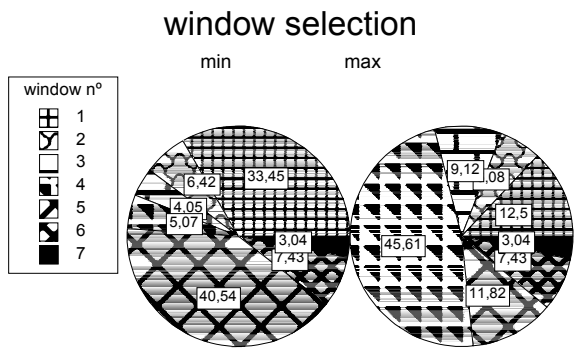

figure 4. Percentage of use of each window

In figures 5.2, 5.3 a sample of the obtained results with $\&\&_2$, $v_1*pinv(v_2)$ is provided, with the sentence "el golpe de timón fue sobrecogedor".

With the selected error measures ($\&\&_2$, $v_1*pinv(v_2)$) there are not significative differences between choosing min, max or median filtering, evaluated with the number of peaks, but selecting the min{} operator, there are the following advantages:
1. An accurate location of the transitions because they are only detected when there is "really" a change, indicated by all the $D_i$. Using max, mean or median, there is an anticipative detection of peaks, which present a flatter and wider aspect.
2. With the first stage of the filter (mean), we obtain a reduction of the correlated noise. With the second one (min), the impulsive noise is obviously detected but it has lower effect over the spectral variation function than mean, max or median filtering.

**5.1 Energy Change Detector.**

Really, the most important parameter is the error measure, because it defines "what is a transition". One way to evaluate the difference between 2 numbers, is his division. Then, if they are similar the result is nearly one, but if they are different, we obtain a big or small number. To obtain always a big number, we must use:

$$\{(a/b) + (b/a)\}$$

In the vectorial case, we can operate in the same way using the idea of the pseudoinverse of a vector (pinv). Then, $v*pinv(v)=1$.

With this error measure, we obtain a very accurate location of the energy changes (See figure 5.3). This change marks are easily and successfully selected with a simple algorithm.

**5.2 Multilevel Segmentation**

The initial variation function (with error criterion $\&\&_2$) is escaled between these marks, to normalize the maximum local value to one. This normalization is realized in order to avoid the fluctuations of the energy along the sentence to be segmented, and simplify the significative peak selection. It is interesting to observe that the shape of the resultant function has the same peak locations, and it is not necessary to provide an adaptive threshold.

Multilevel segmentation can be easily implemented reducing the threshold.

**ACKNOWLEDGEMENTS**

We thank Dr. Climent Nadeu for his helpful suggestions.

**REFERENCES**
[1] F. Bruguara, R. De Mori, D. Giuliani, M. Omologo "Improved connected digit recognition using Spectral Variation Function". ICSLP 92, pp. 627-630.

[2] G. Flammia, P. Dalsgaard, O. Andersen, B. Lindberg "Segment based variable frame rate speech analysis and recognition using Spectral Variation Function". ICSLP 92, pp.983-986.

[3] J.G. Wilpon, B. H. Juang, L. R. Rabiner "An investigation on the use of acoustic subword units for automatic speech recognition". ICASSP 87, pp.821-824.

[4] P. Zamperoni "Maps of distances between vectors of rank-ordered local grey values and their application to a model-free texture segmentation". Workshop on nonlinear digital signal processing. pp. 2.2-7.1 to 2.2-7.6 Tampere 1993.

[5] Pitas, Venetsanopoulos "Nonlinear digital filters: Principles and applications. Ed. Kluwer 1991

[6] M. Sanchez, M. Faúndez, F. Vallverdú "Codificador de voz a baja velocidad con segmentatión fonética". URSI-95, Valladolid.

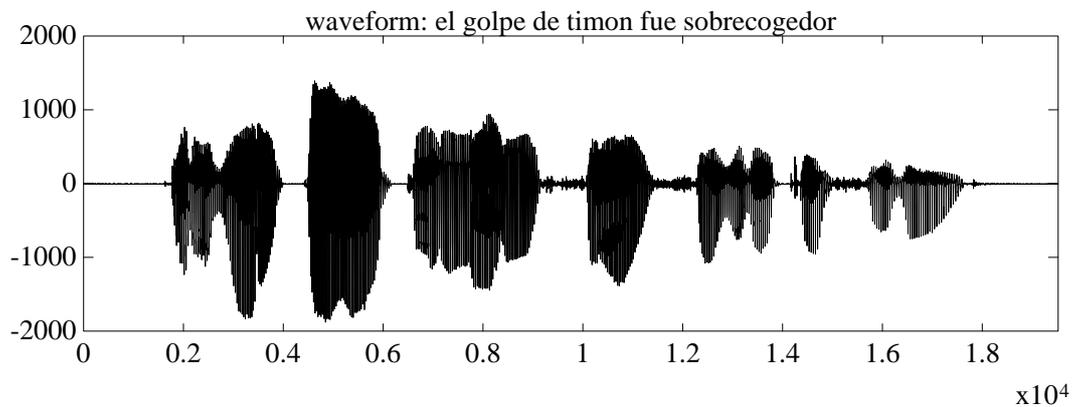

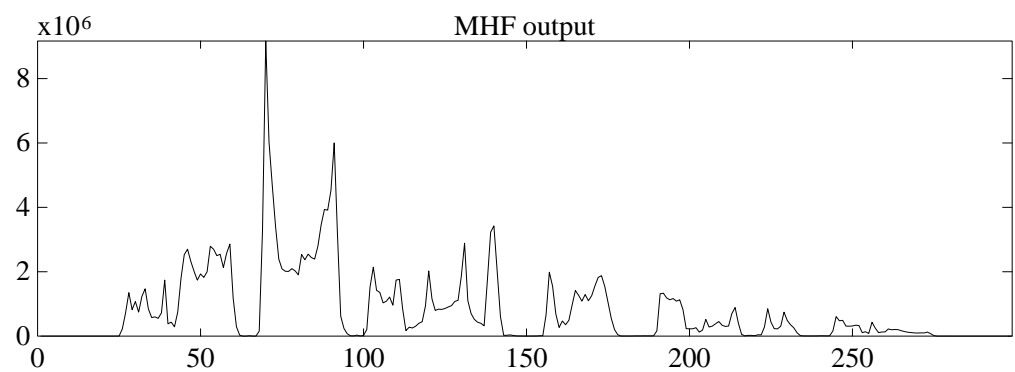

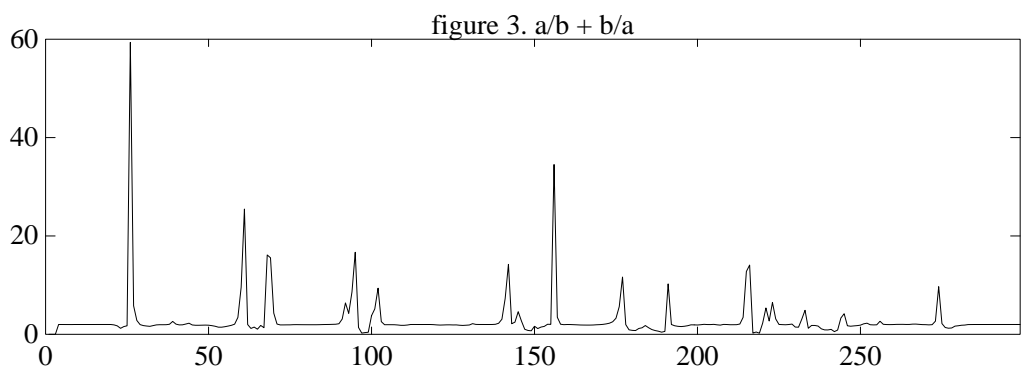

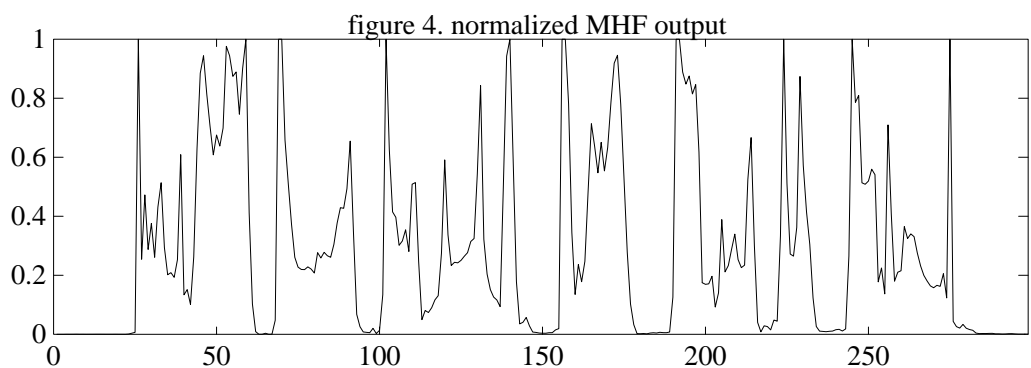

Figure 6. Waveform and outputs of the MHF variants.